# A Critical UV Legacy: A Hubble Roadmap for HWO Science Readiness

Sarah Peacock (*UMBC/NASA/GSFC*), Aiden S. Zelakewicz *(Cornell/CSI),* Lisa Kaltenegger *(Cornell/CSI),* Breanna A. Binder *(CalPoly Pomona),* José A. Caballero (*Centro de Astrobiologia, CSIC-INTA),* Lía Corrales *(University of Michigan),* Kevin France *(LASP/CU-Boulder),* Cynthia Froning *(SwRI),* Eric Mamajek *(JPL),* Seth Redfield *(Wesleyan University),* Tyler Richey-Yowell *(McDonald Observatory),* Keighley Rockcliffe *(UMBC/NASA/GSFC),* Edward Schwieterman *(UC Riverside),* Riccardo Spinelli *(Osservatorio Astronomico di Palermo, INAF),* Noah Tuchow *(University of Arizona),* David Wilson *(LASP/CU-Boulder),* Allison Youngblood *(NASA/GSFC)*

## Executive Summary:

The Habitable Worlds Observatory (HWO) will provide the first opportunity to directly image and spectrally characterize terrestrial exoplanets in the habitable zones of nearby stars. Maximizing its scientific return requires a comprehensive understanding of the high-energy radiation environments of target stars, which shape planetary atmospheres and govern the production, destruction, and detectability of biosignatures.

Ultraviolet (UV) radiation plays a particularly critical role in atmospheric chemistry. Far-ultraviolet (FUV) and near-ultraviolet (NUV) photons regulate key photochemical pathways, influence ozone stability, and drive the formation of prebiotic molecules. However, the majority of high-priority HWO target stars lack high-quality UV observations. Existing datasets are sparse, heterogeneous, or limited by calibration uncertainties, and no comparable UV observatory is expected for at least 5–10 years (with UVEX offering more limited spectral resolution, wavelength coverage, and sensitivity).

The Hubble Space Telescope (HST) remains the only observatory capable of acquiring high-resolution FUV and NUV spectra for these targets over the next 10-15 years. We therefore advocate for a coordinated HST program to systematically obtain UV spectra of high-priority HWO targets, ideally in conjunction with X-ray observations. This effort is essential for enabling accurate target prioritization, constraining stellar radiation environments, and ensuring robust interpretation of future HWO observations.

## Key Science Questions Enabled by HST:

A central question for exoplanet science in the coming decade is: *what radiation environments do habitable-zone planets actually experience?* Stellar UV emission, particularly in the FUV and NUV, governs the photochemistry and thermal structure of planetary atmospheres. For K and M dwarfs, which dominate the nearby stellar population, much of this emission originates in the chromosphere and transition region rather than the photosphere, making it poorly captured by standard stellar atmosphere models [1]. Moreover, stellar UV output varies significantly with magnetic activity, rotation, and age, introducing substantial star-to-star diversity even within a single spectral class [e.g., 2,3,4]. High-resolution UV spectroscopy with HST is uniquely required to resolve key diagnostic lines and disentangle these emission components.

These radiation environments directly impact the interpretation of biosignatures. UV photons drive both the production and destruction of atmospheric species such as $O_2$, $O_3$, $CH_4$, and $CO_2$, and can generate abiotic oxygen and ozone under certain conditions [e.g.,5,6,7,8,9,10,11,12,13]. Without empirical constraints on stellar UV spectra, photochemical models remain underdetermined, and degeneracies in atmospheric retrievals can negatively influence the interpretation of HWO observations. HST provides the necessary empirical inputs to anchor these models and reduce systematic uncertainties.

A related challenge is determining the full high-energy (X-ray-UV) radiation environment of nearby stars. X-ray and extreme ultraviolet (EUV) radiation play a central role in shaping planetary atmospheres by driving ionization, heating, and atmospheric escape. EUV photons, in particular, often dominate the energy budget of the upper atmosphere and set mass loss rates, while also influencing the chemical pathways that regulate biosignature gases [e.g.,14,15,16,17,18]. Because EUV emission cannot currently be observed directly, it must be inferred from X-ray and UV data. In practice, FUV emission lines serve as essential proxies, enabling either (1) broadband EUV flux estimates through empirical scaling relations [e.g., 19,20] or (2) full spectral reconstructions guided by UV diagnostics [e.g., 21,22,23]. Without HST spectroscopy, EUV estimates remain highly uncertain, propagating directly into atmospheric and habitability models.

Finally, stellar UV characterization is itself a prerequisite for identifying optimal HWO targets [1]. Current target lists (The “ExEP List” or TSS25) [24,25] are based primarily on geometric detectability and stellar properties, but do not yet incorporate detailed high-energy radiation environments. Some nominally favorable targets may prove too active for reliable biosignature interpretation, while more quiescent systems may emerge as higher priority. Furthermore, to predict the yield of exoplanets that HWO will observe in the near UV, knowledge of host star UV flux is required. A comprehensive UV dataset is therefore essential not only for interpreting HWO observations, but also for defining the target sample itself.

## The Current UV Data Gap for HWO Targets:

A comprehensive understanding of stellar radiation environments is currently limited by the availability of observational data [1]. The majority of high-priority targets identified for HWO lack modern, high-quality ultraviolet spectroscopy. As shown in Figure 1, less than half of the current high-priority targets have been (or are scheduled to be) observed with HST in both the FUV and NUV bands. While earlier missions, such as IUE, provided partial coverage, these data are often limited by flux calibration uncertainties and scattered light contamination, reducing their utility for precision exoplanet applications. Photometric measurements from GALEX are similarly incomplete and lack the spectral resolution required to constrain stellar radiation fields.

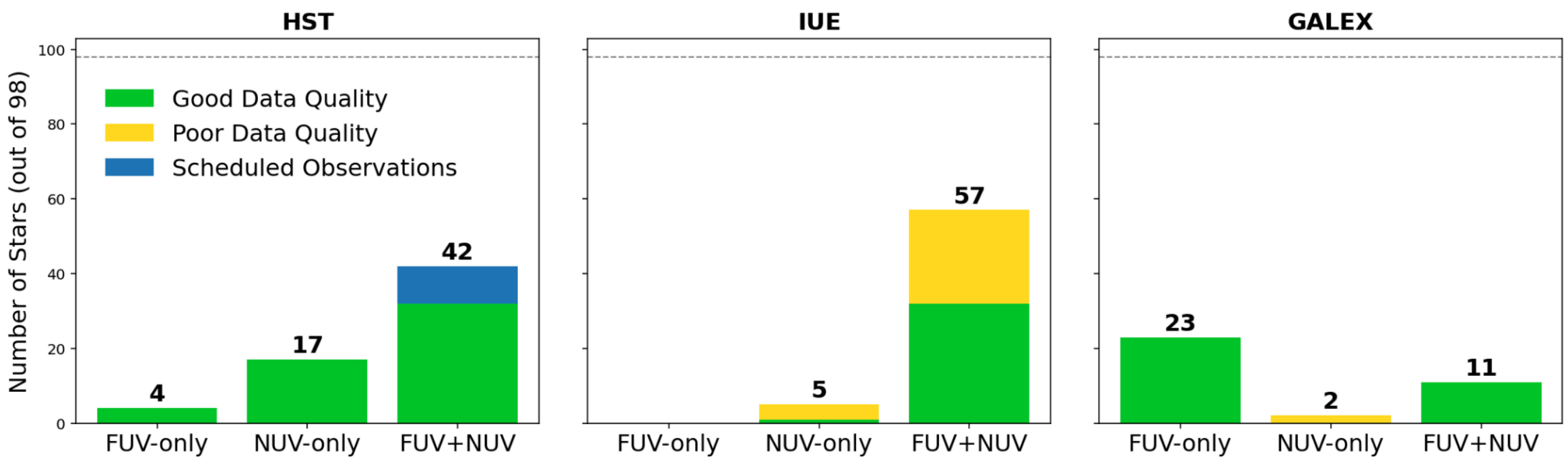

**Figure 1:** *UV observational completeness for high-priority HWO target stars (ExEP Tier A+B). Less than half of targets have complete UV coverage from HST, and only a few have partial FUV or NUV-only spectra. Legacy data from IUE and photometric measurements from GALEX are not sufficient substitutes due to calibration limitations and lack of spectral resolution. This incomplete coverage represents a major source of uncertainty in both HWO target prioritization and atmospheric modeling.*

This incomplete and heterogeneous dataset introduces systematic uncertainties at every stage of HWO science, from exposure time calculations to atmospheric retrievals. Critically, the absence of UV spectra is not randomly distributed: stars with existing data are often biased toward higher activity levels or known exoplanet hosts, leaving a large fraction of quiescent, potentially favorable targets uncharacterized [26]. As a result, current target lists may be systematically skewed, and optimal HWO targets may remain unidentified.

While proxy-based approaches are often used to approximate stellar radiation environments, they break down in the high-energy regime where chromospheric and coronal emission varies significantly even among stars with similar fundamental parameters. As illustrated in Figure 2, G-type stars within the HWO target sample exhibit substantial diversity in their UV and X-ray emission, despite having broadly similar effective temperatures and luminosities. These differences arise from variations in stellar activity, rotation, and age, and directly impact the shape and intensity of the high-energy spectrum. Notably, even the Sun cannot serve as a reliable template for other solar analogs, as adopting solar or template-based spectra can introduce order-of-magnitude discrepancies in key wavelength regions that control atmospheric photochemistry and escape (Figure 2).

## Required Instrument Capabilities and Operations:

Addressing these science questions requires continued access to high-resolution FUV and NUV spectroscopy with HST, specifically through the COS (110-320 nm, R=3000-20000) and STIS (115-318 nm, R=1000-115000) MAMAs. These instruments provide the spectral resolution and sensitivity needed to resolve chromospheric and transition region emission lines, reconstruct intrinsic stellar fluxes, particularly for the key Lyα line, and measure continuum emission relevant for exposure time calculations. Current and upcoming CubeSat and SmallSat missions (e.g., CUTE, SPARCS, MAUVE) lack the necessary effective area and spectroscopic resolution. The NASA Explorer-class mission UVEX (UltraViolet EXplorer) is planned to have a GO program that will enable pointed FUV spectroscopy (R>1000, 115-265 nm) and multi-epoch UV imaging, but will not fully replace the HST capabilities summarized above that are best tailored for the host star characterization.

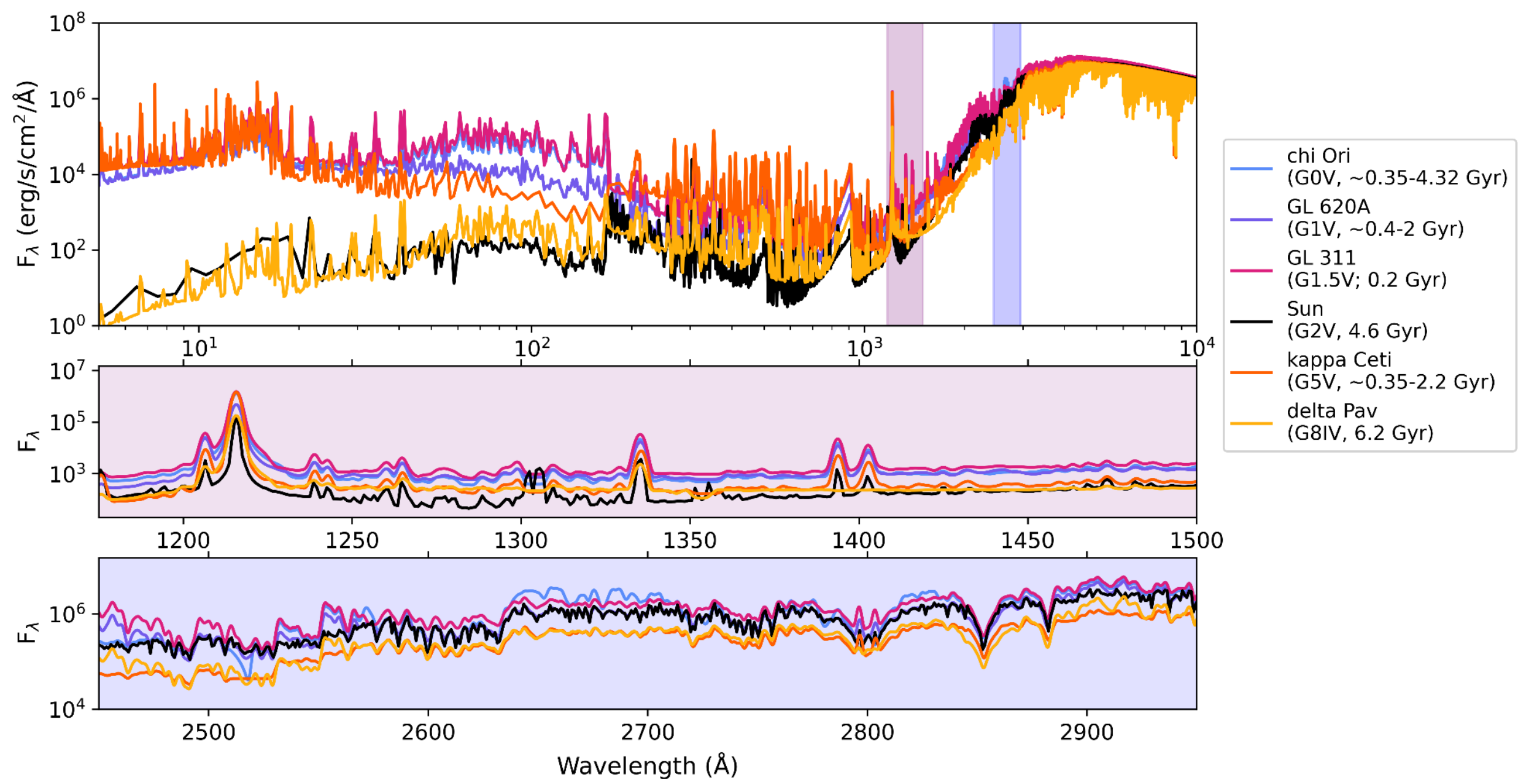

**Figure 2:** *Panchromatic spectral energy distributions for representative G-type stars in the HWO target sample with existing multiwavelength data. Despite similar stellar types, the high-energy emission varies significantly between stars, reflecting differences in magnetic activity, age, and rotation. A solar spectrum (black) illustrates that even the Sun is not representative of other solar analogs in the high-energy regime. These variations lead to large discrepancies in the radiation environments experienced by orbiting planets, demonstrating that proxy spectra are insufficient and that direct UV observations of each target are required.*

Accurate absolute flux calibration is critical, as UV flux levels directly inform HWO coronagraph exposure times and yield estimates. Stability across multiple epochs is also required to characterize variability and ensure consistent comparisons across targets. Broad wavelength coverage spanning both the FUV and NUV is necessary to capture the full range of diagnostic features, including Lyman-alpha, Mg II h&k, and other key emission lines.

Observationally, multi-epoch strategies are essential to capture variability due to flares, rotational modulation, and stellar activity cycles. Coordinated or contemporaneous observations with X-ray facilities such as Chandra X-ray Observatory and XMM-Newton are strongly recommended to enable full characterization of the XUV spectrum. Deep integrations will be required for intrinsically faint targets, particularly late-type stars that are likely among the most promising HWO hosts.

### Preparing for HWO: The Role of HST

HST plays a foundational role in preparing for HWO in several key ways [27]:

- **Coronagraph performance models:** Provide NUV fluxes for HWO target stars, enabling accurate exposure time calculations, integration time estimates, and survey yield predictions.
- **Atmospheric modeling & biosignature interpretation:** Supply realistic stellar UV spectra that set the photochemical and climate boundary conditions, improving predictions of atmospheric composition and spectral signatures.

- **EUV reconstruction:** Enable indirect reconstruction of the unobservable EUV spectrum using FUV diagnostics (in combination with X-ray data), critical for modeling atmospheric heating, escape, and long-term habitability.
- **Reducing uncertainties in stellar radiation environments:** Anchor the stellar input to exoplanet studies, enabling propagation of uncertainties in UV fluxes into retrievals.
- **Constraining LISM absorption:** Constrain local interstellar medium absorption in Ly-alpha and Mg II h&k lines, which is essential for recovering intrinsic stellar UV fluxes.

## Recommended Large-Scale Initiative:

We recommend the development of an HST Stellar UV Legacy Survey to obtain FUV and NUV spectra for all high-priority HWO target stars [24,25] (including completing and updating existing datasets) in order to create a definitive UV spectral library for nearby stars that can serve as a foundation for HWO science and broader astrophysical applications. A key component of this effort should be coordination with X-ray observatories to obtain contemporaneous or near-contemporaneous observations, enabling full XUV characterization. Time-domain coverage should also be incorporated through multi-epoch observations, allowing characterization of stellar variability on timescales ranging from flares to activity cycles.

This initiative would produce a high-impact, legacy dataset with broad utility across stellar astrophysics and exoplanet science while directly enabling the core science goals of HWO. The observations are required for each individual target, as the HWO sample spans a wide range of spectral types, ages, and magnetic activity levels that are not captured by stellar analogs.

## Conclusions:

HST represents a unique and time-limited opportunity to obtain the UV data required to fully realize the scientific potential of HWO. No current or planned facility can replicate the sensitivity, spectral resolution, and wavelength coverage in the FUV and NUV over the coming decade. Without a concerted effort to characterize the UV radiation environments of nearby stars, the interpretation of HWO observations will be fundamentally limited by stellar uncertainties. A strategic HST program to obtain these data will establish the empirical foundation needed for accurate target selection, robust atmospheric modeling, and confident biosignature detection. HST is an essential precursor to HWO: investing in UV observations now will ensure that HWO delivers on its promise to identify and characterize habitable worlds beyond our solar system.